\documentstyle[graphicx,psfig]{mn}

\def\spose#1{\hbox to 0pt{#1\hss}}

\def\multleft#1{\hbox to size{\vbox {\halign {\lft{##}\cr #1}}\hfill}\par}
\def\multright#1{\hbox to size{\vbox {\halign {\rt{##}\cr #1}}\hfill}\par}

\def\boxit#1{\vbox{\hrule\hbox{\vrule\kern3pt\vbox{\kern3pt
          #1 \kern3pt}\kern3pt\vrule}\hrule}}

\def\cm{{\rm\thinspace cm}}

\def\erg{{\rm\thinspace erg}}

\def\keV{{\rm\thinspace keV}}
\def\km{{\rm\thinspace km}}
\def\kpc{{\rm\thinspace kpc}}

\def\Mpc{{\rm\thinspace Mpc}}

\def\s{{\rm\thinspace s}}
\def\yr{{\rm\thinspace yr}}

\def\pcmcu{\hbox{$\cm^{-3}\,$}}

\def\ergpcmcu{\hbox{$\erg\cm^{-3}\,$}}
\def\ergpspcmsq{\hbox{$\erg\cm^{-2}\s^{-1}\,$}}

\def\ergps{\hbox{$\erg\s^{-1}\,$}}

\def\kmps{\hbox{$\km\s^{-1}\,$}}

\def\pcmsq{\hbox{$\cm^{-2}\,$}}

\def\kmpspMpc{\hbox{$\kmps\Mpc^{-1}$}}

\title[The Interaction of 3C~401 with the Surrounding ICM]{The Interaction of 3C~401 with the Surrounding Intra-Cluster Medium}

\author[C.~S.~Reynolds, L.W.Brenneman, J.T.Stocke]{
\parbox{15cm}{
Christopher~S.~Reynolds$^1$,
Laura~W.~Brenneman$^1$ and
John~T.~Stocke$^2$}\\
$^1$Dept.\ of Astronomy, University of Maryland, College Park, MD 20742, USA.\\
$^2$Dept. of Astrophysics and Planetary Science, University of Colorado, Boulder, CO~80309,USA.}

\date{In press}
\pagerange{\pageref{firstpage}--\pageref{lastpage}}
\pubyear{2004}

\begin{document}
\label{firstpage}
\maketitle

\begin{abstract}
  We present an observation of the radio-galaxy 3C~401 and the surrounding
  intracluster medium (ICM) of its host galaxy cluster by the {\it
    Chandra X-ray Observatory}.  This luminous radio-galaxy is notable
  in that it has characteristics intermediate between the FRI and FRII
  morphologies.  We clearly detect point-like emission coincident with
  the radio-core of 3C~401, although the spatial resolution of even
  {\it Chandra} is only 2\,kpc at the distance of 3C~401 ($z=0.201$)
  and so the possibility remains that this is a dense (and rapidly
  cooling) thermal gaseous core in the center of the ICM atmosphere.
  Strong departures from spherical symmetry in the central 10--20\,kpc
  of the ICM clearly suggest interaction between the ICM and the
  radio-lobes of 3C~401.  A central X-ray bar probably results from
  the evacuation of two ICM cavities by the expanding radio lobes.
  Beyond these central regions, the cluster possesses a flatter
  profile than many clusters of comparable mass suggesting the
  importance of ICM heating and entropy injection by 3C~401.  We
  detect an interesting cross-like structure in the ICM on 100\,kpc
  scales.  We speculate that this could be a radio-galaxy induced
  disturbance corresponding to a time when 3C~401 was substantially
  more powerful.  A particularly exciting possibility is that this
  cross-like structure corresponds to a large scale global g-mode
  oscillation excited by a past outburst of 3C~401.
\end{abstract}


\section{Introduction}

The classic ``cooling flow problem'' (Fabian 1994), the discrepancy
between the observed radiative cooling rate of the intracluster medium
(ICM) in rich galaxy clusters and the amount of cold gas or star
formation actually observed, has obvious connections to galaxy
formation.  One can rephrase the cooling flow issue into the question
``Why are the cD galaxies in rich clusters not still in the process of
forming?''  It seems clear that some agent must be heating the ICM
core of rich clusters in order to (on average) balance the radiative
cooling, and an obvious candidate for that agent is AGN activity.  In
other words, radio-galaxy/ICM interactions may well be the agent that
terminates the formation of the most massive galaxies (Benson et al.
2003; Binney 2004).

While a few examples of radio-galaxy/ICM interactions were known in
the {\it Einstein} and {\it ROSAT} days, it took the superior imaging
capability of {\it Chandra} to reveal the full complexity and ubiquity
of this phenomenon.  For example, Perseus~A (Fabian et al. 2000,
2003), Hydra~A (McNamara et al. 2000; David et al.  2001; Nulsen et
al.  2002), Abell~2052 (Blanton et al. 2001), and Cygnus~A (Smith et
al.  2002) all show well defined cavities in the X-ray emitting gas
which are coincident with the current radio lobes of the central radio
galaxy.  In these sources, it is clear that the radio lobes have
displaced the X-ray emitting gas producing the observed X-ray/radio
anti-coincidence.  B\^irzan et al. (2004) have used an analysis of 16
clusters with cavities to find a significant correlation between the
mechanical energy required to create the cavities and the radio power
level of the central radio source; the existence of this correlation
is strong evidence that the radio source is responsible for evacuating
these cavities. Additionally, in at least 50\% of the cases studied,
the mechanical energy estimated to be present in the cavities is
sufficient to offset the estimated cooling of the cluster core.  {\it
  Chandra} has also revealed the presence of ``ghost'' cavities, i.e.,
X-ray cavities that are {\it not} coincident with the active radio
lobes.  In the B\^irzan et al. (2004) study the ``ghost cavities''
have significantly larger estimated mechanical energies than cavities
containing current-outburst radio source lobes.  Examples include the
outer cavity of Perseus~A (Fabian et al.  2000, 2003), Abell~2597
(McNamara et al.  2001), NGC~4636 (Jones et al. 2002), and Abell~4059
(Heinz et al.  2002; Choi et al. 2004).  In these sources, it is
believed that the cavities are associated with old radio lobes
(related to previous cycles of AGN activity).  The low-frequency
(74\,MHz) synchrotron radio emission expected within this scenario has
been observed from the ghost cavity of Perseus-A (Fabian et al. 2002).

If one is interested in assessing the wider importance of
radio-galaxy/ICM interactions (e.g., to galaxy formation processes),
it is of obvious use to extend the study to powerful radio-galaxies
beyond our local universe.  In this paper, we describe an observation
of the moderately powerful ($P_{\rm 20cm}= 5\times10^{26}\,{\rm
  W}\,{\rm Hz}^{-1}$) radio-galaxy 3C~401 ($z=0.201$) by the {\it
  Chandra X-ray Observatory}.  Despite having the radio power of an
FR~II, this source has been classified as intermediate between an FR~I
and FR~II morphologically (Harvanek \& Stocke 2000), raising the
possibility that it might be an example of a fading radio source.
Specifically, 3C~401 contains no highly concentrated ``hot spots'' at
the leading edges of its two lobes and the brightest portion of its
extended structure is a luminous jet in the southern lobe.  These
characteristics are more similar to those of typical FR~Is than
FR~IIs.  3C~401 is also much broader compared to its length than a
typical FR~II; thus, the nickname given to sources with this
morphology: ``fat doubles''. See Harvanek \& Stocke (2002) for an
identification and discussion of other ``fat doubles'', including
Hercules A.  3C~401 is surrounded by a cluster of galaxies with
optical galaxy density (with the galaxy-galaxy two point correlation
function B$_{gg} \approx$ 1100 Mpc$^{1.77}$; Harvanek et al. 2001),
equivalent to Abell richness class I -- II.  Harvanek et al. (2001)
found that ``fat doubles'' are found exclusively in clusters of
galaxies at intermediate redshift.

The goal of the {\it Chandra} observation reported in this paper was
to search for and characterize any interaction between this
radio-galaxy and the ICM of its host galaxy cluster.  Clear signatures
of this interaction were found.  Section~2 describes the observation
and our data reduction.  Both our spectral and imaging results are
presented in Section~3, and placed into a wider context in Section~4.
Our conclusions are drawn in Section~5.  Assuming the {\it WMAP}
Cosmology (flat universe with $\Omega_{\Lambda}=0.73$;
$H_0=71\kmpspMpc$; Spergel et al. 2003) gives a luminosity distance of
976\,Mpc, an angular size distance of 678\,Mpc, and a look-back time
of 2.42\,Gyr\footnote{N.Wright Cosmology Calculator via the NASA
  Extragalactic Database.}.  At this distance, 1\,arcsec subtends a
linear distance of 3.38\,kpc.

\section{Chandra observations and data reduction}

The 3C~401 system was observed by {\it Chandra} on the 20-Sept-2002
(22.7\,ks exposure; obs-id 3083) and the 21-Sept-2002 (24.9\,ks
exposure; obs-id 4370).  The core of 3C~401 was centered 1\,arcmin
from the aim point on the S3 back-illuminated ACIS chip (7) such that
the core of its cluster could be imaged entirely by the S3 chip.
These observations were taken in imaging-mode, i.e., the diffraction
gratings are not placed in the X-ray path.  The two level-2 events
files for these two observations were merged using the {\tt merge\_all}
script, and then filtered so as to include only events with ASCA
grades 0,2,3,4, and 6.  Exposure corrected images and spectra were
created following the standard process outlined in the {\it Chandra
  Interactive Analysis of Observations} (CIAO) analysis threads
\footnote{http://cxc.harvard.edu/ciao/threads/}.  All ACIS data
reduction was performed using the CIAO version 3.0, and spectral
analysis was performed using version 11.3 of the {\sc xspec} fitting
package.

\section{Results}

\subsection{Large scale X-ray image}
\label{fig:imaging}

\begin{figure}
\centerline{
\psfig{figure=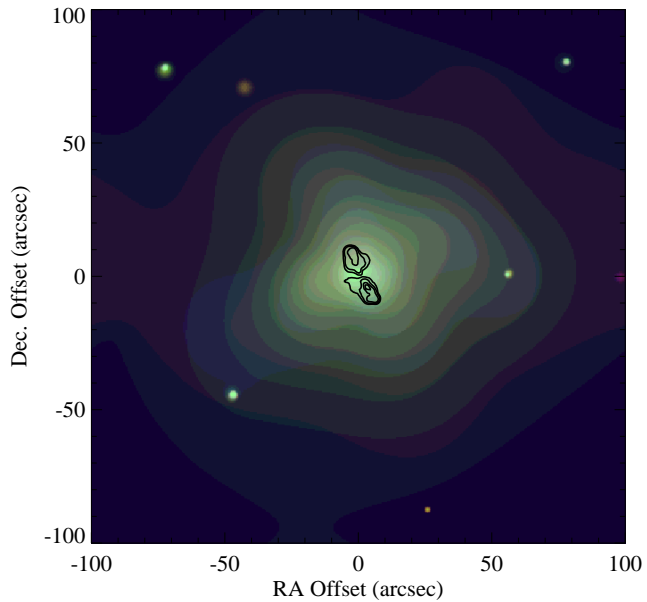,width=0.4\textwidth}
}
\vspace*{1cm}
\caption{Adaptively smoothed three-colour Chandra/ACIS-S image of 
  the 3C~401 cluster overlaid with contours of 20cm radio-emission
  from MERLIN.  Note the cross-like structure displayed by the ICM.
  See text for a discussion of how this image was constructed and the
  robustness of the cross-like structure.  North is up in this image.}
\label{fig:large_image}
\end{figure}

Figure~\ref{fig:large_image} shows an adaptively smoothed 3-color {\it
  Chandra} image of the general region surrounding 3C~401, overlaid
with contours of 20\,cm emission from MERLIN\footnote{The radio
  observation was performed 5-May-1993, and the reduced radio data
  were obtained from the NASA Extragalactic database.}.  To produce
this X-ray image, we first extracted and exposure-corrected images
from the cleaned filtered level-2 events file in three bands;
0.3--0.8\,keV (soft band), 0.8--2\,keV (medium band), and 2--10\,keV
(hard band).  The total image (i.e., the sum of the three bands) was
adaptively smoothed using a $3-4\sigma$ smoothing kernel, and the
resulting map of smoothing lengths was then applied to each of the
images in the three bands separately.  The three images are then
overlaid to form the 3-color image, with red, green and blue denoting
the soft, medium and hard band images respectively.  At the distance
of 3C401, this image covers a square region 680\,kpc on a side.

As can be seen from Fig.~\ref{fig:large_image}, we clearly detect an
extended ICM halo centered on 3C~401, although we fail to detect any
spatial variations in the hardness of the emission.  Interestingly,
the ICM appears to possess a cross-like structure, with four almost
perpendicular spurs extending in the NNE/ESE/SSW/WNW directions.  The
NNE/SSW axis is exactly that defined by the radio-jets of 3C~401.
Given that this is an adaptively smoothed image, one might be
concerned that these spurs extend approximately in the direction of
the point sources also seen in the field.  However, precisely the same
spur pattern is observed if one first removes the point sources prior
to producing the adaptive smoothing map.  One might also be concerned
about the validity of the exposure map used to produce these images.
However, the spurs are seen in both exposure corrected and
non-exposure corrected images, and the amplitude of variations across
the spurs (30--50\%) is substantially greater than can reasonably be
explained by any exposure map effect.  Thus, we conclude that these
spurs are real.  We return to this issue in Section~\ref{sec:cluster}
where we further verify the reality of these spurs through surface
brightness profiles derived from unsmoothed data.

We now proceed to discuss the core of this system and, in particular,
evidence for radio-galaxy/ICM interaction.

\subsection{ICM/Radio-Galaxy Interaction}
\label{sec:rg_icm_int}

\begin{figure*}
\hbox{
\hspace{1cm}
\psfig{figure=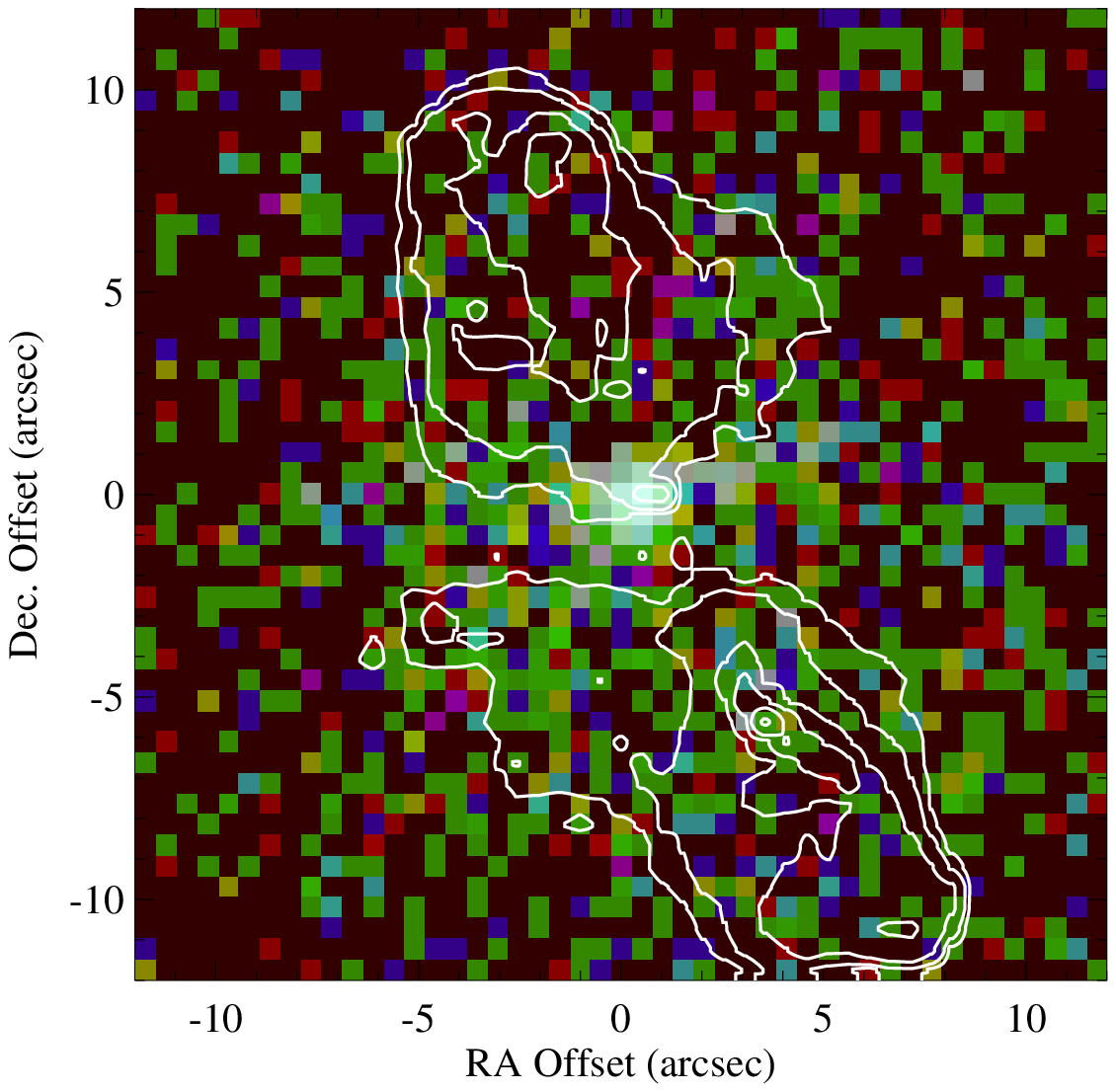,width=0.4\textwidth}
\hspace{2cm}
\psfig{figure=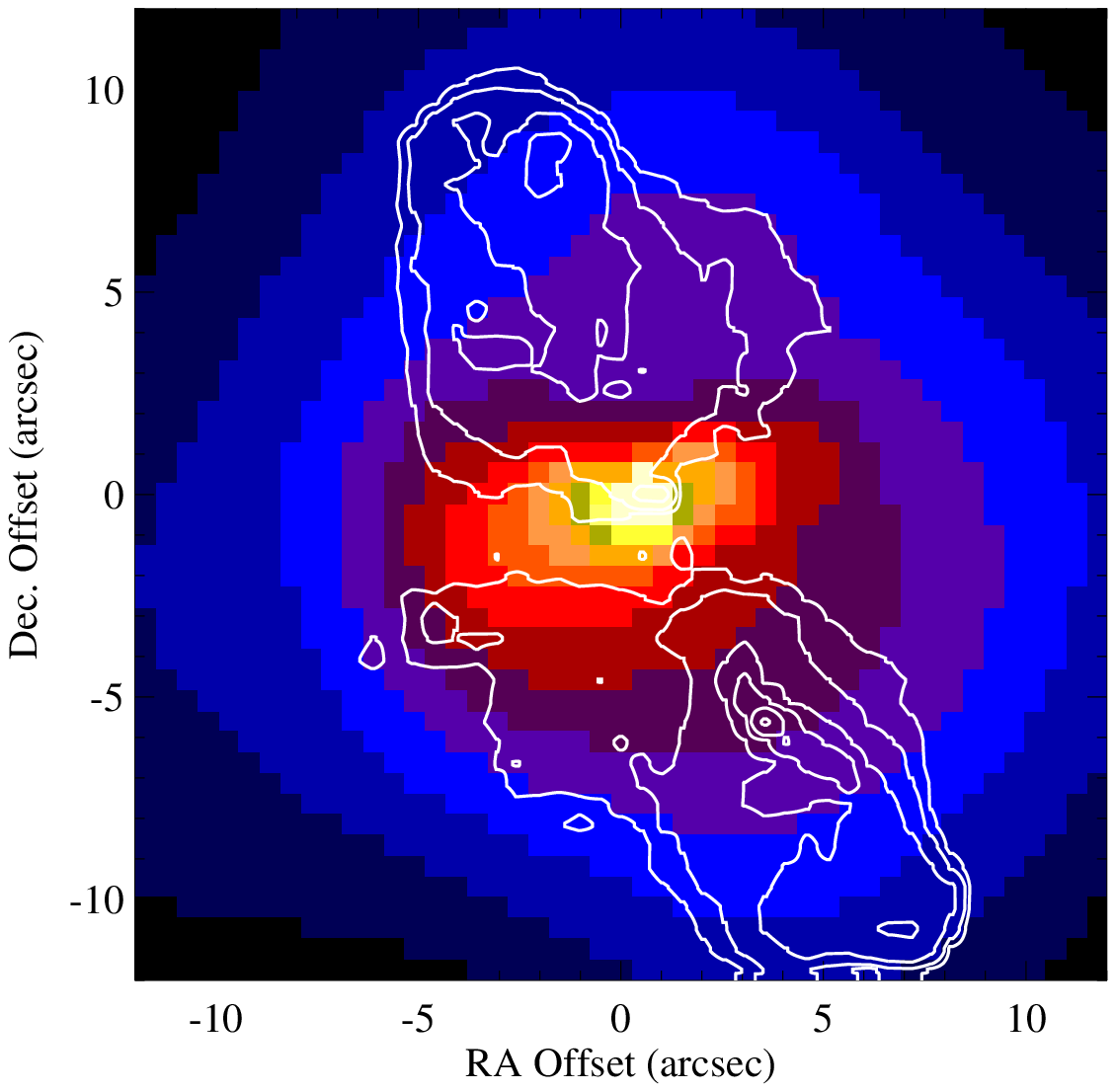,width=0.4\textwidth}
}
\vspace{1cm}
\caption{{\it Left panel: }Three-colour raw photon image of the central 
  $24\times 24\arcsec^2$ regions of the 3C~401 system overlaid with
  contours of 20cm radio emission from MERLIN.  The nucleus of 3C~401
  lies at the center of this image and is coincident with the bright
  point X-ray source. {\it Right panel: }Adaptively smoothed total
  intensity image of this region.  The anti-coincidence between the
  radio lobes and the X-ray emission is evident and results in the
  formation of a central X-ray bar. }
\label{fig:small_image}
\end{figure*}

In Fig.~\ref{fig:small_image}, we show both a raw 3-color X-ray image
and an adaptively smoothed total intensity X-ray image of the central
regions of the 3C~401 system.  As before, the X-ray images have been
overlaid with contours of 20\,cm radio emission from MERLIN.  As is
apparent in the raw image, there is a luminous point-like X-ray source
at the location of the 3C~401 radio core (see Section~\ref{sec:agn}
for more discussion of this source).  Although the photon statistics
are not good, the ICM emission is clearly not spherically-symmetric.
In particular, there is a notable deficit of counts $3-5\arcsec$ north
of the central source compared with the same distance east or west of
the central source.  This deficit is clearly shown in the smoothed
total intensity map, which also reveals a similar deficit in the
region 3--5\,arcsec south of the central source.  In other words,
there is a strong nuclear bar apparent, with a position angle
orthogonal to the radio axis.

Furthermore, careful inspection of the raw image reveals hints of a
cone-like structure bounding the northern radio-lobe.  However, this
structure does not appear to survive the adaptive smoothing process
and so its reality cannot be confirmed.

\begin{figure*}
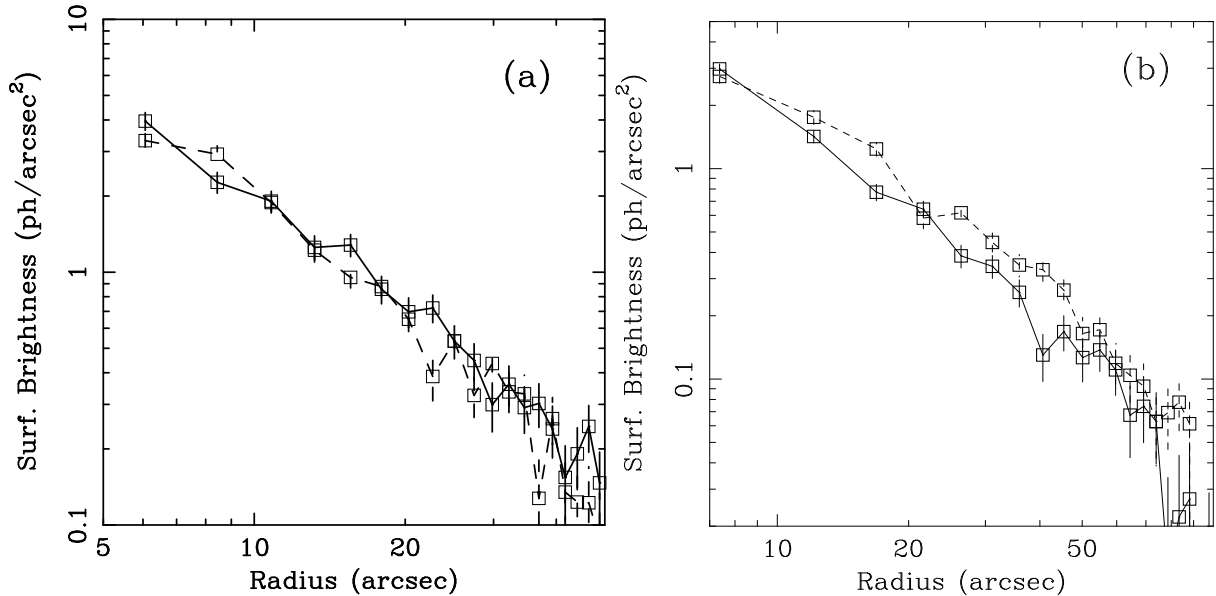

\centerline{
\psfig{figure=f3a.ps,width=0.45\textwidth,angle=270}
\psfig{figure=f3b.ps,width=0.45\textwidth,angle=270}
}
\caption{(a) X-ray surface brightness profiles for a back-to-back pair 
  of quadrants that include the radio lobes (dashed line) and the
  orthogonal pair of quadrants that avoid the radio lobes (solid
  line). (b) X-ray surface brightness profile analysis using eight
  45\,degree sectors centered on 3C~401.  The solid line shows the
  profile averaged over the four alternating sectors that avoid the
  spurs of the cross-like structure seen in Fig.~1.  The dashed line
  shows the profile for the complementary sectors that are coincident
  with the spurs.  The clear excess seen in this latter profile
  verifies that the cross-like structure is not an artifact of the
  adaptive smoothing procedure.}
\label{fig:radial_profile}
\end{figure*}

We further investigate these features by forming radial surface
brightness profiles.  Figure~\ref{fig:radial_profile}a shows the
surface brightness profile within quadrants that are centered on the
radio-axis of 3C~401 (the on-axis profile; red line), as well as the
profile within the orthogonal quadrants (the off-axis profile; green
line).  The nuclear bar-like structure gives rise to a factor of two
enhancement in the off-axis surface brightness at the smallest radii
probed in our surface brightness analysis (7 pixels or 3.5\,arcsecs).
There are other subtle differences between the two radial profiles (in
the form of relative enhancements in the off-axis profile at
15\,arcsecs and 22\,arcsecs.  These differences, while statistically
significant with respect to Poisson noise, have no obvious
corresponding features in either the raw or smoothed the X-ray image.
Deeper X-ray images will be required to assess the nature of these
particular deviations from circular symmetry.

\subsection{The nuclear region}
\label{sec:agn}

A point source is clearly
detected coincident with the radio core of 3C~401.  Although the
spatial resolution of {\it Chandra}/ACIS is approximately 2\,kpc at
the distance of 3C~401, it seems likely that we are detecting the
X-ray emission from the active galactic nucleus itself.  Using a
smaller than usual extraction region (1\,arcsec radius) in order to
separate this source from the surrounding ICM emission, we detect 385
photons in this source, approximately 10--20\% of which are accounted
for by a reasonable extrapolation of the underlying ICM surface
brightness profile.  We have searched for but fail to detect temporal
variability of this central source.  However, our limits on temporal
variability are weak due to the limited photon statistics ---
inspection of the light curve suggests that the source has not varied
by more than 30\% on the timescale of the total observation length
($1.2\times 10^5\s$), and has not varied by more than a factor of two
on a timescale of $10^4\s$.  

Using our small extraction radius, we have extracted a time-average
spectrum of this central source.  We produce a ``background'' spectrum
from a region offset by 3\,arcsec in the western direction.
Therefore, in an attempt to isolate the pure nuclear spectrum, the
background spectrum also includes thermal emission from the ICM
immediately neighboring the nuclear source.  The source spectrum was
then binned so as to contain at least 15 photons per energy bin, hence
allowing the use of $\chi^2$ statistics.

The 0.5--6\,keV nuclear spectrum can be adequately fit with either a
power-law or thermal plasma model modified by the effect of Galactic
absorption ($N_H=6.8\times 10^{20}\pcmsq$) and absorption by
contaminants on the ACIS filter (modeled with the {\tt acisabs}
model).  Note that there are inadequate numbers of photons above
6\,keV to allow meaningful extension of these fits to higher energies.
Best fit parameters for the power-law fits are; photon index
$\Gamma=1.89^{+0.27}_{-0.21}$, intrinsic (redshifted) absorption
$N_{Hz}<1.0\times 10^{21}\pcmsq$, observed 0.5-10\,keV flux
$F_{0.5-10}=5.5\times 10^{-14}\ergpspcmsq$, and intrinsic (unabsorbed)
0.5--10\,keV luminosity $L_{0.5-10}=7.1\times 10^{42}\ergps$, with a
goodness of fit parameter $\chi^2/{\rm dof}=18.4/19$.    

The power-law model will be appropriate if we are, indeed, observing
AGN emission.  However, the possibility remains that this central
source could be compact (unresolved) core of thermal emission from the
hot gas halo.  Indeed, fitting such a model to the data using the {\tt
  mekal} model encoded in the {\sc xspec} spectral fitting package
(version 11.3.0; Mewe, Gronenschild \& van~den~Oord 1985; Mewe, Lemen
\& van~den~Oord 1986; Kaastra 1992; Liedahl, Osterheld \& Goldstein
1995) results in a fit which is almost as good as (and statistically
indistinguishable from) the power-law model.  Best fit parameters
for this thermal plasma model are; plasma temperature
$kT=4.9^{+2.5}_{-1.8}\keV$, plasma abundance $Z<2.6\,Z_\odot$,
intrinsic (redshifted absorption) $N_{Hz}<5\times 10^{20}\pcmsq$,
observed 0.5-10\,keV flux $F_{0.5-10}=5.1\times 10^{-14}\ergpspcmsq$,
and intrinsic (unabsorbed) 0.5--10\,keV luminosity
$L_{0.5-10}=6.7\times 10^{42}\ergps$, with a goodness of fit parameter
$\chi^2/{\rm dof}=21.3/18$.  The emission measure suggested by this
spectral fit is $EM=4.1\times 10^{65}\pcmcu$.  If we make the simple
and conservative approximation that this thermal plasma uniformly
fills a sphere of the size of our extraction region, this emission
measure implies a plasma density of $n>0.25\pcmcu$, and a
bremsstrahlung cooling timescale of $t_{\rm brems}<1-2\times 10^8\yr$.
Thus, if this central source is indeed a dense gaseous core, it
possesses a cooling timescale which is much shorter than any realistic
age of this cluster.

\subsection{The surrounding cluster}
\label{sec:cluster}

Finally, we discuss the gross properties of the ICM emission.  We
extracted a spectrum from a region centered on 3C~401 with an
extraction radius of 32\,arcsec (108\,kpc).  A background spectrum was
formed from blank sky observations.  After background subtraction,
this spectrum possesses 3360 photons.  After binning the spectrum to
at least 15 photons per energy bin, we fit the 0.5--8\,keV spectral
data with a thermal plasma model modified for the effects of Galactic
and instrumental absorption (as discussed in Section~\ref{sec:agn}).
The best fit parameters are: plasma temperature $kT=2.9\pm 0.3\keV$,
plasma abundance $Z=0.43^{+0.20}_{-0.16}\,Z_\odot$, observed
0.5-10\,keV flux $F_{0.5-10}=3.7\times 10^{-13}\ergpspcmsq$, and
intrinsic (unabsorbed) 0.5--10\,keV luminosity $L_{0.5-10}=5.6\times
10^{43}\ergps$, with a goodness of fit parameter $\chi^2/{\rm
  dof}=116/120$.  The emission measure (EM) for this plasma is
measured to be $EM=4.3\times 10^{66}\pcmcu$.

It is also interesting to examine the spatial structure of the
cluster.  By combining the two radial surface brightness profiles
discussed in Section~\ref{sec:rg_icm_int} (also
Fig.~\ref{fig:radial_profile}) we have produced the
azimuthally-averaged surface brightness profile.  To this profile, we
fit a standard $\beta$-model in which the surface brightness is given
by
\begin{equation}
\label{eq:beta}
S(r)=\frac{S_0}{\left[1+(r/r_0)\right]^{3\beta-1/2}}.
\end{equation}
In a free fit, we obtain $r_0=36\pm 8\kpc$ and
$\beta=0.46^{+0.03}_{-0.02}$ ($\chi^2=11.4$ for 17 degrees of
freedom).  This is a rather flat density profile and small core radius
for such a cluster to possess.  If we fix $\beta=0.67$, the canonical
value measured in rich clusters of galaxies, the fit is substantially
worse ($\chi^2=80$ for 18 degrees of freedom) due to a underprediction
of surface brightness at the smallest radii ($<5\arcsec$) and largest
radii ($>30\arcsec$) and an overprediction of the surface brightness
in the range 10-20\,arcsec.

Inverting eqn.~\ref{eq:beta} to give the radial dependence of the
volume emissivity, and normalizing using the value of $EM$
determined from the spectral fit allows us to deduce the approximate
radial run of density $n(r)$ and pressure $p(r)$ (assuming that the
ICM is isothermal with $kT=2.9\keV$).  The result of this exercise is
\begin{eqnarray}
n(r)&=&1.8\times 10^{-2}\left[1+\left(\frac{r}{36\kpc}\right)^2\right]^{-3\beta/2}\pcmcu,\\
p(r)&=&8.3\times 10^{-11}\left[1+\left(\frac{r}{36\kpc}\right)^2\right]^{-3\beta/2}\ergpcmcu.
\end{eqnarray}
Knowing the absolute values of density and pressure is essential for
examining the energetics of the radio-galaxy interaction, a topic that
we shall address in the next Section.

We finish our discussion of surface brightness profiles by returning
to the issue of the large scale spurs or cross-like structure noted in
the adaptively smoothed surface brightness map (Fig.~1).  Due to the
four-fold symmetry, we do not expect these features to be clearly
revealed in the quadrant analysis presented above.  To examine a
putative structure with four-fold symmetry requires that we perform
the following tailored surface brightness profile analysis.  We divide
the image plane into eight equal and non-overlapping 45\,degree
sectors centered on the core of 3C~401.  This ``wheel'' of sectors is
aligned so that the radio-axis of 3C~401 lies on the mid-line of one
back-to-back pair of sectors.  We then form an average radial surface
brightness profile using the four (alternating) sectors that lie
either along or perpendicular to the radio axis
(Fig.~\ref{radial_profile}b; dashed line).  The adaptively smoothed
image suggests that these sectors should coincide with the four spurs
of the cross-like structure.  We compare this with the average radial
surface brightness profile using the other four sectors which,
according to the adaptively smoothed image, should lie in the
inter-spur gaps (Fig.~\ref{radial_profile}b; solid line).  This
analysis reveals a clear excess of counts in the on-spur surface
brightness profile as compared with the off-spur case at radii of
20-50\,arcsec (70--170\,kpc).  Thus, we conclude that the cross-like
structure is not an artefact of the adaptive smoothing.

\section{Discussion}
\label{sec:discussion}

Our {\it Chandra} observations have revealed three striking aspects of
the 3C~401 system.  Firstly, there is clear evidence for an ongoing
interaction between 3C~401 and the ICM of its surrounding galaxy
cluster.  The most obvious manifestation of this interaction is the
nuclear bar (with radius of $\sim 10\arcsec$/$35\kpc$) that lies
orthogonal to the axis of the MERLIN radio lobes.  Guided by
observations of radio-galaxy/ICM interactions in the local universe
($z<0.1$), it seems very likely that this structure results from the
formation of ICM cavities by the expanding radio-lobes.  The detailed
anti-coincidence between the X-ray surface brightness and the radio
surface brightness seen in 3C~401 supports this hypothesis.  Using our
estimates of the ICM pressure, we can make a crude estimate for the
mechanical power required to inflate these radio lobes.  The total
energy required to inflate the two lobes is $E\sim 1.5\times
10^{59}\erg$, where we have approximated each lobe as a sphere with
radius $5\arcsec$ and have taken the required energy to be $E\sim 2pV$
where $V$ is the volume of the lobe.  The lifetime of the source is
likely to be of the order of the ICM sound crossing time of one radio
lobe, $t\sim 1.7\times 10^{15}\s$.  Thus, the time-averaged power
required to inflate these radio-lobes against the pressure of the ICM
is $P=E/t\sim 1\times 10^{44}\ergps$.  This is very close to the
measured X-ray luminosity of the ICM showing that the radio galaxy can
have a major impact on the energetics of the ICM in this source if the
mechanical energy can be thermalized efficiently.  Taking the
spatially integrated 1.4\,GHz flux to be $4.7\times 10^{-23}\ergps$
(Kellerman, Pauliny-Toth \& Williams 1969), we estimate the radio
power to be $\nu L_\nu\approx 7\times 10^{42}\ergps$.  This places the
3C~401 cluster near the correlation between the mechanical energy
inferred to be inside these cavities and the current radio power level
(B\^irzan et al. 2004). Consistency with this correlation means that
the current radio source contains sufficient mechanical energy to
create these cavities (assuming the theoretical expectation of a
1--10\% efficiency for converting the total radio source mechanical
energy into radio frequency luminosity; Bicknell et al. 1997).

Secondly, we have noted a larger scale cross-like structure extending
to distances of $\sim 50\arcsec$ ($170\kpc$) from the centre of the
cluster and also aligned with the radio-axis of 3C~401.  While the
reality of this feature appears to be robust, its interpretation is
not clear.  The coincidence between the orientation of this structure
and the radio axis of 3C~401 suggests that this might also be due to
radio plasma interaction with the ICM, although the possibility
remains that the cross-like structure is caused by unrelated dynamical
processes (e.g., subcluster mergers).  If it is indeed due to
interaction with 3C401, two possibilities arise.  If this ICM
structure is caused by two pairs of ``ghost cavities'', then they are
amongst the largest known.  Using the same assumptions as in the
paragraph above to assess the energetics of these ghost cavities, we
estimate that 3C401 had to have a period about 300\,Myr ago in which its
mechanical power was $2-5\times 10^{44}\ergps$ (i.e., a factor of a
few greater than the present).  While this explanation for the ICM
cross has the appeal that ghost cavities are structures that are {\it
  known} to exist in some clusters, it does not naturally explain the
four-fold symmetry of this structure (the two pairs of ghost cavities
would have to lie at roughly the same distance from 3C~401 and have
axes that are perpendicular).

This leads us to speculate that the ICM cross is actually due to a
high amplitude global oscillation mode (most likely a low-$l$ internal
gravity mode) excited by a previous outburst from 3C401.  The theory
of such oscillations has been developed by Balbus \& Soker (1990) and
Lufkin, Balbus and Hawley (1995), although these authors envisage the
excitation of internal g-modes through a resonant interaction with
orbiting galaxies, not through an explosive central event.  A detailed
theoretical investigation of this possibility, including predicted
maps of ICM surface brightness and temperature for different modes, is
beyond the scope of this paper.  At this stage, we note that the
oscillation period of such a mode will be a factor of a few longer
than the sound crossing time of the region, and the energy of the mode
will be of the same order as that estimated above for the ghost cavity
scenario.  However, it is likely that only a modest fraction of the
total energy from the radio-galaxy outburst would end up in such a
mode, with p-modes likely carrying away the majority of the energy of
the initial blast.  Hence, within this scenario, 3C~401 would likely
have exceeded a mechanical luminosity of $10^{45}\ergps$ during its
past phase of activity.  Deeper imaging of this field (most likely
with XMM-{\it Newton} will be required to study the ICM cross
structure in more detail and distinguish between (or disprove) the
ghost cavity and global mode scenarios.

The final striking aspect of this system is the unusual surface
brightness distribution of the cluster --- we measure a core radius of
$r_0=36\kpc$ and $\beta=0.46$, substantially flatter than the typical
$\beta=0.67$ found in many clusters of a comparable or greater mass.
However, this is quite similar to the value of $\beta$ found in low
mass clusters and groups (Osmond \& Ponman 2004 and references
therein), a result that is taken as evidence for the enhanced
importance of excess entropy in these low mass systems.  Whether the
3C~401 cluster really is anomalous in having a flat profile for its
mass requires further study with deeper X-ray imaging.  In particular,
one may be concerned that we are not obtaining a true measure of the
value of $\beta$ given that the central region of the cluster is
morphologically complex and that we cannot constrain the ICM surface
brightness profile beyond about 170\,kpc.  If the flat profile is
confirmed, it is tempting to interpret this as signs of particular
strong ICM heating and entropy injection, as might be expected for a
cluster whose ICM is still in the process of forming.  Indeed, we
proposed to observe this radio galaxy/ICM system with {\it Chandra}
because previous work (Harvanek et~al. 2001; Harvanek \& Stocke 2002)
presented significant evidence linking intermediate FR~I/FR~II radio
galaxies to the formation of a dense ICM in the cluster which
surrounds them.

\section{Conclusions}

Chandra imaging spectroscopy of the intermediate FRI/FRII radio-galaxy
3C~401 reveals clear signs of an interaction between the radio-galaxy
and the ICM of its surrounding cluster.  Although we do not see well
defined ICM cavities (most likely due to the limited number of photons
in our image), the flattening of the central X-ray isophotes strongly
suggests that the radio-lobes are indeed evacuating cavities within
the ICM with a radius of 15--20\,kpc.  There are also deviations from
spherical symmetry on much larger spatial scales (100\,kpc or more)
that reveal themselves as a cross-like structure in the low surface
brightness regions of the ICM.  It is tantalizing to speculate that a
previous, much more powerful period of activity from 3C~401 created
these large scale disturbances.  While this large scale cross-like
pattern might be caused by a pair of ghost cavities related to
previous activity, the symmetry of this structure leads us to
speculate that the ICM atmosphere is executing large-amplitude,
low-$l$ g-mode oscillations.  Deeper imaging of this field (either
with a very long {\it Chandra} observation or a moderately deep {\it
  XMM-Newton} observation) is required to further probe these
possibilities.

The ICM possesses a temperature of $kT\approx 2.9\keV$ and a
0.5--10\,keV luminosity of $L_{0.5-10}=5.6\times 10^{43}\ergps$
placing it firmly on the standard L-T relationship of Horner (2002).
This cluster also possesses an X-ray luminosity typical for its
cluster richness (B$_{gg}$ value; Yee \& Ellingson 2003), although its
T$_x$ is nearly a factor of two below that expected for its richness
(Yee \& Ellingson 2003). In addition, its surface brightness profile
is unusually flat.  When parameterized by a standard King-type model,
the ICM possesses $\beta=0.46$, much flatter than the normal
$\beta=0.67$ found in clusters of comparable or greater mass.  These
unusual ICM features may all be due to the unusually large radio power
level of the central galaxy for a cluster with a dense ICM. However,
because this analysis uses only a few thousand X-ray photons total, we
worry that the ICM surface brightness distribution may not have been
measured robustly as yet.

This system is ripe for deeper X-ray imaging spectroscopy, either with
a long {\it Chandra} stare or a moderately long {\it XMM-Newton}
observation.  With such data, we will be able to search for the
well-defined ICM cavities that probably encase the currently active
radio-lobes, study the cross-like structure on 170\,kpc scales, and
constrain the ICM surface brightness profile significantly beyond
100\,kpc.  New data of this type will take studies of the current
radio-galaxy/ICM interaction as well as the ICM thermodynamics of the
3C~401 system to the next level.

\section*{Acknowledgments}

We thank Steve Balbus, Mitch Begelman and Andy Fabian for stimulating
discussions through the course of this work.  We acknowledge support
from {\it Chandra} Cycle-3 Guest Observer Program under grant
G02-3143X and the National Science Foundation under grant AST0205990.


\begin{thebibliography}{}

\bibitem[]{} Balbus S.A., Soker N., 1990, ApJ, 357, 353
\bibitem[]{} Benson A.J., Bower R.G., Frenk C.S., Lacey C.G., Baugh C.M., Cole S., 2003, ApJ, 599, 38
\bibitem[]{} Bicknell G.V., Dopita M.A., O'Dea C.P., 1997, ApJ, 485, 112
\bibitem[]{} Binney J., 2004, Phil. Trans. Roy. Soc., in press (astroph/0407238)
\bibitem[]{} Birzan L., Rafferty D.A., McNamara B.R., Wise M.W., Nulsen P.E.J., 2004, ApJ, 607, 800
\bibitem[]{} Blanton E.L., Sarazin C.L., McNamara B.R., Wise M.W., 2001, ApJ, 558, L15
\bibitem[]{} Choi Y.Y., Reynolds C.S., Heinz S., Rosenberg J.L., Perlman E.S., Yang J., 2004, ApJ, 606, 185
\bibitem[]{} David L.P. et al., 2001, ApJ, 557, 546
\bibitem[]{} Fabian A.C., 1994, ARA\&A, 32, 277 
\bibitem[]{} Fabian A.C. et al., 2000, MNRAS, 318, L65
\bibitem[]{} Fabian A.C. et al., 2002, MNRAS, 331, L35
\bibitem[]{} Fabian A.C. et al., 2003, MNRAS, 344, L43
\bibitem[]{} Harvanek M., Stocke J.T., 2002, AJ, 124, 1239
\bibitem[]{} Harvanek M., Ellingson E., Stocke J.T., Rhee G., 2001, AJ, 122, 2874 
\bibitem[]{} Heinz S., Choi Y.Y., Reynolds C.S., Begelman M.C., 2002, ApJ, 569, L79
\bibitem[]{} Horner D., 2002, PhD thesis (univ. of Maryland).
\bibitem[]{} Jones C et al., 2002, ApJ, 567, L115
\bibitem[]{} Kaastra J.S., 1992, An X-ray Spectral Code for
Optically-Thin Plasmas (Internal SRON-Leiden Report, updated version
2.0)
\bibitem[]{} Kellerman K.I., Pauliny-Toth I.I.K., Williams P.J.S., ApJ, 157, 1
\bibitem[]{} Liedahl D.A., Osterheld A.L., Goldstein W.H., 1995, ApJL, 438, 115
\bibitem[]{} Lufkin E.A., Balbus S.A., Hawley J.F., 1995, 446, 529
\bibitem[]{} McNamara B.R. et al., 2000, ApJ, 534, L135
\bibitem[]{} McNamara B.R. et al., 2001, ApJ, 562, L149
\bibitem[]{} Mewe R., Gronenschild E.H.B.M., van~den~Oord G.H.J., 1985,
A\&AS, 62, 197
\bibitem[]{} Mewe R., Lemen J.R., van~den~Oord G.H.J., 1986, A\&AS, 65, 511
\bibitem[]{} Nulsen P.E.J. et al., 2002, ApJ, 568, 163
\bibitem[]{} Osmond J.P.F., Ponman T.J., 2004, MNRAS, 350, 1511
\bibitem[]{} Smith D.A., Wilson A.S., Arnaud K.A., Terashima Y., Young A.J., 2002, ApJ, 565, 195
\bibitem[]{} Spergel D.N., et al., 2003, ApJS, 148, 175
\bibitem[]{} Yee H.K.C., Ellington E., 2003, ApJ, 585, 215

\end{thebibliography}
\end{document}